\documentclass{ws-ijmpe}
\usepackage{amsmath}
\usepackage{amssymb}
\usepackage{graphicx}
\usepackage{bm}
\newcommand{\be}{\begin{equation}}
\newcommand{\ee}{\end{equation}}

\newcommand{\bea}{\begin{eqnarray}}
\newcommand{\eea}{\end{eqnarray}}

\begin{document}

\markboth{X. Vi\~nas, M. Centelles, M. Warda}{Semiclassical 
description of exotic nuclear shapes}

\catchline{}{}{}{}{}


\title{SEMICLASSICAL DESCRIPTION OF EXOTIC NUCLEAR SHAPES}
\author{X. VI\~NAS\textsuperscript{a}, M. CENTELLES\textsuperscript{a} and 
M. WARDA\textsuperscript{a,b}}

\address{
\textsuperscript{a} Departament d'Estructura i Constituents de la 
Mat\`eria
and Institut de Ci\`encies del Cosmos, \\
Facultat de F\'{\i}sica, Universitat de Barcelona,
Diagonal {\sl 647}, {\sl 08028} Barcelona, Spain\\
\textsuperscript{b} Katedra Fizyki Teoretycznej, Uniwersytet Marii 
Curie--Sk\l odowskiej,\\
        ul. Radziszewskiego 10, 20-031 Lublin, Poland}  

\maketitle

\begin{history}
\received{(received date)}
\revised{(revised date)}
\end{history}

\begin{abstract}
Exotic nuclear structures such as bubbles and tori are analyzed through
semiclassical extended Thomas-Fermi calculations with the Skyrme force
SkM$^*$. The variational equations for neutron and proton densities
are solved fully self-consistently in spherical (bubbles) and
cylindrical (tori) symmetries. The possible existence of bubble
configurations in some astrophysical scenarios is discussed. The
stability of toroidal structures against change of quadrupole moment is
studied. A global minimum of the energy  is found in heavy
toroidal nuclei.
\end{abstract}

\section{Introduction}

The idea that stable or metastable nuclei could exist in the form of
spherical bubbles was suggested many years ago.\cite{Wil46,Whe50,Sie67}
In the seventies Wong\cite{Won73} investigated the stability of exotic
shapes, namely spherical bubbles and tori, in the framework of the
liquid drop model (LDM)\cite{LDM} with Strutinsky shell
corrections.\cite{Str} Wong found that within the LDM bubble and
toroidal structures were unstable against some deformations, but they
could be stabilized with the help of shell corrections. He addressed his
study to known nuclei near the the $\beta$-stability valley. More
recently, Dietrich and Pomorski\cite{Die97} have carried out a similar
analysis for bubble nuclei with large mass numbers. They have found
stability islands for spherical bubbles in the range of mass numbers
$A$ between 450 and 3000. Possible large bubble structures have also
been analyzed within the mean-field approximation with the
Hartree-Fock-Bogoliubov (HFB) theory and the D1S Gogny
force.\cite{Dec99,Dec03} 

Semiclassical methods of Thomas-Fermi (TF) type are a powerful tool for the
study of nuclear properties that smoothly vary with the number of particles $A$.
Similarly to the LDM the TF methods are free of shell corrections, but they
describe the nuclear surface in a consistent way.\cite{Bra74} As is known, the
self-consistent density profiles at the pure TF level end at the classical
turning point. This deficiency can be cured by using the extended Thomas-Fermi
(ETF) method\cite{Bra85,Cen90} which adds to the pure TF approach $\hbar^2$ (and
$\hbar^4$) corrections coming from the non-commutativity between the position
and momentum operators. The ETF approach allows to obtain self-consistent
density profiles with a fall-off similar to the one obtained with the mean field
Hartree-Fock (HF) or HFB calculations.

The TF and ETF approaches are free of shell effects and replace the
wave functions by the proton and neutron local densities as the main
variables. This fact makes these approaches particularly useful for
dealing with calculations with a very large number of particles or
that involve the continuum (as the description of hot nuclei) where
complete HF calculations may be extremely complicated.

Our aim in this paper is to study with some detail the ETF predictions
for large spherical bubbles and toroidal distributions. It is a step
towards a more complete description of these structures, that will
imply to compute the shell energy with the same effective interaction
used to calculate the ETF energy. In the second section we briefly
outline the basic theory. In the third section we present our results
for spherical bubbles and discuss the existence of two different
density distributions with almost the same energy per particle for the
same nucleus. Some attention is paid to the possibility that bubble
structures can appear in model astrophysical scenarios. In the fourth
section we analyze toroidal distributions in deformed ETF calculations
constrained by a negative mass quadrupole moment. Our conclusions and
outlook are given in the last section.

\section{Basic theory}

In this work we will use for our semiclassical study of bubble and  toroidal
structures the zero-range SkM$^*$ interaction\cite{SkM*} which is known to give
realistic fission barriers in actinide nuclei.\cite{Bra85} This seems a
reasonable choice for describing nuclear structures containing several hundreds
of nucleons.

The Skyrme energy density functional can be written as:
\begin{equation}
{\cal E}={\cal E}_{kin} + {\cal E}_{zr} + {\cal E}_{den} + {\cal E}_{fin}
+ {\cal E}_{so} + {\cal E}_{Coul},
\label{eq1} \end{equation}
where ${\cal E}_{kin}$ is the kinetic energy including the 
density-dependent effective mass, ${\cal E}_{zr}$ and ${\cal E}_{den}$ 
are the zero-range central and density-dependent terms which are functions 
of the neutron and proton densities $\rho_n$ and $\rho_p$  
only, ${\cal E}_{fin}$ depends on
$\mbox{\boldmath $\nabla$}\rho_n$ and $\mbox{\boldmath $\nabla$}\rho_p$ 
and simulates the finite range of the effective nucleon-nucleon 
interaction, and finally ${\cal E}_{so}$ and ${\cal E}_{Coul}$ are the
contributions to the energy density coming from the spin-orbit and
Coulomb interactions, respectively.

In the semiclassical ETF approach up to $\hbar^2$ order, the kinetic and 
spin-orbit energy densities are expressed as functionals of the local 
density as follows:
\begin{eqnarray}
{\cal E}_{kin} & = & \sum_q \frac{\hbar^2 \tau_q}{2 m_q^*} =
\sum_q \frac{\hbar^2}{2 m}f_q \tau_q =
\sum_q  \frac{\hbar^2}{2 m}
\bigg\{ \frac{3}{5}(3\pi^2)^{2/3}f_{q}\rho_{q}^{5/3} + 
\frac{1}{36}f_{q}\frac{(\mbox{\boldmath $\nabla$}
 \rho_q)^2}{\rho_q} 
\nonumber \\[3mm]
& & \mbox{}
- \frac{1}{3} \mbox{\boldmath $\nabla$}f_q 
\cdot \mbox{\boldmath $\nabla$}\rho_{q}
- \frac{1}{12} \rho_q \frac{(\mbox{\boldmath $\nabla$}f_{q})^2}{f_{q}} +
\frac{1}{2} \frac{\rho_{q}}{f_q} \big(\frac{2 m}{\hbar^2}\big)^2
({\bf W}_{q})^2 \bigg\}, 
\label{eq2} \end{eqnarray}   
\begin{equation}
{\cal E}_{so} =
- \frac{2 m}{\hbar^2}\sum_q \frac{\rho_{q}}{f_q} ({\bf W}_{q})^2, 
\label{eq3} \end{equation}
where $q=n,p$ for neutrons and protons, respectively. The ratio of the
bare nucleon mass to the effective mass $f_q$ and the spin-orbit
potential ${\bf W}_{q}$ are the standard expressions associated to the
Skyrme energy density functional.\cite{Bra85}

The Coulomb energy density reads
\begin{equation}
{\cal E}_{Coul} = \frac{1}{2} \rho(\vec{r}) V_{Coul}(\vec{r})
- \frac{3}{4} e^2 {\big(\frac{3}{\pi} \big)}^{1/3} \rho_p^{4/3}(\vec{r}),
\label{eq4} \end{equation}
where we obtain the Coulomb potential $V_{Coul}(\vec{r})$
by solving the discrete Poisson equation 
$\Delta V_{Coul}(\vec{r}) = -4 \pi \rho_p(\vec{r})$ using Gaussian 
elimination.

From the energy density (\ref{eq1}) the variational Euler-Lagrange (EL) 
equations with the constraint on given number of neutrons and protons read
\begin{equation}
\frac{\delta}{\delta \rho_q}  \int d \vec{r} ( {\cal E} - \mu_n \rho_n -
\mu_p \rho_p )
 = \frac{\partial {\cal E}}{\partial \rho_q} - \mbox{\boldmath $\nabla$}
\frac{\partial {\cal E}}{\partial \mbox{\boldmath $\nabla$} \rho_q}
+ \Delta \frac{\partial {\cal E}}{\partial \Delta \rho_q} - \mu_q = 0.
\label{eq5} \end{equation}

The explicit expressions for the EL equations associated to the ETF
energy functional up to $\hbar^2$ order can be found in Ref.\
\cite{Bra85}. We solve the set of two coupled nonlinear second-order
differential equations (\ref{eq5}) by means of the so-called imaginary
time-step method (ITSM).\cite{Dav80} The procedure we follow is
described in detail in Appendix B of Ref.\ \cite{Cen90}. Our solution of
Eqs.\ (\ref{eq5}) is fully self-consistent, with no initial assumption
about the shape of the neutron and proton density profiles which are
obtained by solving Eqs.\ (\ref{eq5}) once two boundary conditions are
imposed. The EL equations of the ETF model have been solved
self-consistently in spherical and cylindrical coordinates using the
ITSM\cite{Dal85}. The deformed ETF model has been successfully applied
to the semiclassical study of fission barriers with
non-relativistic\cite{Bra85,Gar89} and relativistic\cite{Cen93}
interactions. 

\section{Spherical bubbles}

In the left panel of Fig.~1 we display the neutron and proton density
profiles of a nuclear system with $N=560$ neutrons and $Z=240$ protons
as a representative example of bubble nuclei. In this case there exist
two solutions of the EL equations (\ref{eq5}) with very close
energy per nucleon. One of the solutions corresponds to
a sort of hyperheavy nucleus with a reduced but finite central
density, that are called semi-bubbles or unsaturated
nuclei.\cite{Dec99,Dec03} The other solution, a true bubble,
corresponds to a distribution where nucleons are concentrated in a
relatively thin layer around the surface of a sphere. Actually, for a
given atomic number $Z$, there are several values of $N$ for which
both solutions, semi-bubbles and true bubbles, coexist, similarly to
the shape coexistence phenomenon in normal nuclei in the actinide region.
This is illustrated in Table 1 where the energy per nucleon as well as
the neutron and proton chemical potentials are reported for the $Z=240$ isotopic
chain from the neutron to the proton drip lines of the bubble configuration.
From Table 1 we see that semi-bubbles and true bubbles coexist in the range
$N=524$ (proton drip line for semi-bubbles) up to $N$=576. Between $N=524$ and
$N=498$ (proton drip line for true bubbles) only the true bubble solution is
stable against proton emission, although solutions corresponding to unstable
semi-bubbles (with a positive proton chemical potential) can also be found.
Above $N=576$ and up to $N=620$ (neutron drip line for the true bubbles) we only
obtain a true bubble solution.

From Table 1 we also realize that along the $Z=240$ isotopic chain the
ground-state configuration corresponds to the true bubble solution. It
has been found in HF calculations\cite{Dec99,Dec03} that the situation
is reversed for smaller values of $Z$, where the semi-bubble
configuration becomes the ground state. We have checked that this also
occurs in the ETF calculations. The energies per nucleon in
semi-bubbles and bubbles reported in Table 1 are much smaller in
absolute value than in normal nuclei. This is due to the fact that the
binding mechanism of these huge nuclear clusters stems from a delicate
balance between the increase of the surface energy and the reduction
of the Coulomb energy obtained by putting the protons as separated as
possible. 

The neutron and proton single-particle potentials (spp) are displayed on the
right panel of Fig.~1. Roughly, neutron and proton spp follow the corresponding
densities. The neutron spp vanishes at the center of the nucleus for the true
bubble solution while it is considerably reduced with respect to its minimal
value in the semi-bubble case. The proton spp develops strong barriers at the
center and at the outer part of the true bubbles, whereas the semi-bubble only
develops the outer Coulomb barrier as expected.

\begin{figure}
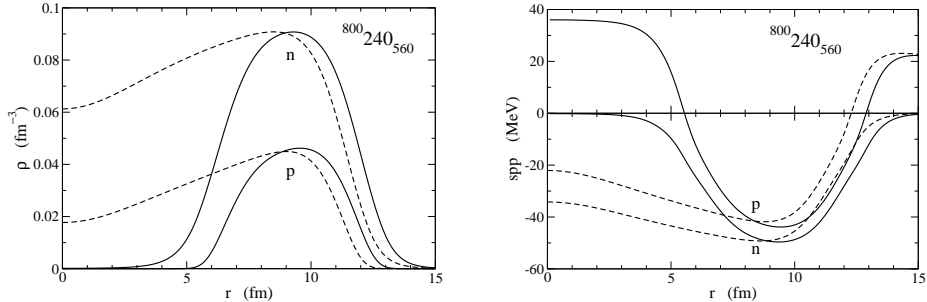

\begin{center}
\psfig{file=kazi_240_N560_bubl_nucl.eps,width=5.8cm,clip=true}
\qquad
\psfig{file=kazi_240_N560_bubl_nucl_SPP.eps,width=5.6cm,clip=true}
\caption{Neutron and proton density profiles (left) and single-particle 
potentials (right) of a true bubble (solid line)
and a semi-bubble (dashed line) with $Z=240$ protons and
$N=560$ neutrons.}  
\end{center}
\end{figure}

\begin{table}
\tbl{The energy per nucleon in MeV as well as neutron and proton
chemical  potentials in MeV of semi-bubbles and true bubbles for $Z=240$
isotopes.}
{\begin{tabular}{rrrrrrrr}
\toprule
      & \multicolumn{3}{c}{semi-bubbles}
      & & \multicolumn{3}{c}{bubbles}  \\
    \cline{2-4}                  \cline{6-8}
$^NZ$ & E/A & $\mu_n$ & $\mu_p$ & & E/A & $\mu_n$ & $\mu_p$ \\
\colrule
%
$^{620}$240 & ---     & ---     & ---     & &$-$4.219 &$-$0.003 &$-$6.409 \\
$^{578}$240 & ---     & ---     & ---     & &$-$4.416 &$-$0.696 &$-$4.480 \\
$^{576}$240 &$-$4.370 &$-$0.576 &$-$2.984 & &$-$4.424 &$-$0.734 &$-$4.380 \\
$^{560}$240 &$-$4.441 &$-$0.898 &$-$2.135 & &$-$4.494 &$-$1.047 &$-$3.558 \\
$^{550}$240 &$-$4.484 &$-$1.115 &$-$1.585 & &$-$4.535 &$-$1.256 &$-$3.025 \\
$^{524}$240 &$-$4.587 &$-$1.731 &$-$0.076 & &$-$4.635 &$-$1.848 &$-$1.569 \\
$^{498}$240 &$-$4.673 &$-$2.428 &   1.543 & &$-$4.719 &$-$2.516 &$-$0.017 \\
\botrule
\end{tabular}}
\end{table}

Heavy-ion collisions at intermediate bombarding energies give strong
evidences for the formation of highly excited nuclei. After some dissipation
due to the pre-equilibrium emitted particles and rotation, the nucleus can
reach an excited equilibrium state with some excitation energy $E^*$. This
excited state can be described by means of a temperature $T$ if the
thermalization time is shorter than the typical decay scale. This state is
metastable with respect to nucleon emission and an artificial pressure
has to be supplied in numerical calculations
for avoiding that nucleons leave the nucleus. A
possible way to do it is to put the nucleus in equilibrium with a
nucleon gas representing the evaporated nucleons. It is assumed that
this artificial pressure on the nucleus will not alter the actual
physical picture of the excited nucleus. As soon as the temperature
is not zero, the continuum states start to be occupied owing to the
thermal Fermi factor. One way of isolating the hot nucleus of the
continuum states was suggested by Bonche, Levit and Vautherin.\cite{BLV}
They found that at a given temperature and chemical
potential, there exist two solutions of the HF equations. One
corresponds to the nucleus in equilibrium with its evaporated gas, and
the other one to the nucleon gas alone. They proposed to study the
nucleus by means of a thermodynamical potential $\tilde{\Omega}$
calculated as the difference of the grand potential $\Omega$ associated
to each one of these two solutions. This subtraction procedure can
also be applied in the semiclassical TF or ETF approaches as shown by
Suraud.\cite{Sur87}

It is worth to point out that when temperature increases shell
corrections become less important and they nearly vanish around $T
\sim 2-3$ MeV.\cite{Bra74} This fact implies that semiclassical
approaches of TF type, which are free of shell effects, are very well
suited for dealing with calculations at high temperature. On the other
hand, HF calculations have shown that there exists a limiting
temperature $T_{lim}$ beyond which normal nuclei become unstable
because of the Coulomb interaction. When temperature grows, the
surface energy decreases faster than the Coulomb contribution and,
consequently, nucleons are driven out of the nucleus for a
sufficiently high temperature.\cite{Sur87}

\begin{figure}
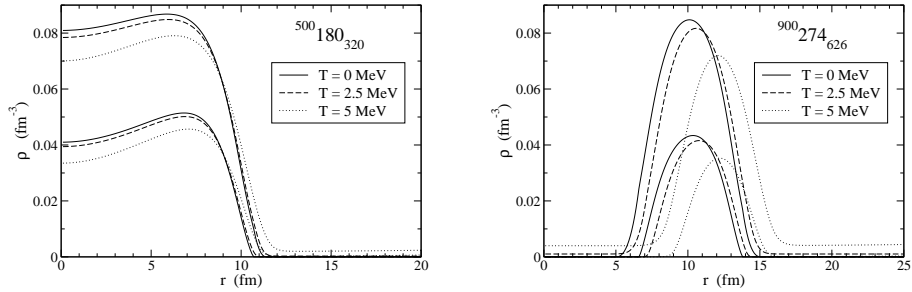

\begin{center}
\psfig{file=kazi_B500newT.eps,width=5.6cm,clip=true}
\qquad
\psfig{file=kazi_B900newT.eps,width=5.6cm,clip=true}
\caption{Neutron and proton density profiles at T=0, 2.5 and 5 MeV
of bubbles containing 500 (left) and 900 (right) nucleons.}
\end{center}
\end{figure}

As far as at zero temperature spherical bubbles become stable against
deformation due to shell effects, it could be expected that spherical bubbles
should disappear around $T \sim 2-3$ MeV when shell effects are washed out.
However, using a simple model it has been shown that thermal effects could
provide an extra stabilization against deformation.\cite{Mor97} We have
performed thermal TF calculations with the semi-bubble $^{500}$180 and with the
true bubble $^{900}$274 finding in both cases a limiting temperature of about 5
MeV. This limiting temperature in bubbles is smaller than the one found in
normal nuclei, which lies in the range $T_{lim} \sim 8-10$ MeV\@. This fact can
be expected due to the large number of protons in the bubbles considered here
which favor the Coulomb instability when temperature grows. At TF level the
excitation energy per nucleon at the limiting temperature is $3.40$ MeV$/A$ for
the semi-bubble $^{500}$180 and $2.06$ MeV$/A$ for the true bubble $^{900}$274.
This is in agreement with the fact that in normal nuclei the limiting excitation
energy per nucleon decreases with increasing mass number. 
The fact that huge spherical bubbles cannot sustain a large amount of
excitation energy has also been qualitatively discussed in Ref.\
\cite{Die97}. In Fig. 2 the
self-consistent TF neutron and proton density profiles corresponding to the
liquid plus gas phase\cite{BLV,Sur87} at $T =0, 2.5$, and 5 MeV are displayed.
When temperature increases nucleons are pushed from the interior to the outer
part of the system increasing the rms radius and the surface diffuseness of the
density distributions.

The fact that isolated spherical bubbles that are stable against neutron
and proton emission contain a large number of nucleons makes it
difficult to obtain such nuclear clusters through heavy-ion collisions.
As far as exotic nuclear shapes such as rods, pasta, etc.\ have been
conjectured to occur in neutron stars\cite{Pet95}, it makes some sense
the question whether exotic bubbles {\it as considered here} can exist
in some astrophysical scenario. To investigate this we make use here of
the simple approach developed in Ref. \cite{Sil02}. It is assumed that in
nuclear matter at subnuclear densities the nuclei are located in a
lattice which we treat in the spherical Wigner-Seitz (WS) approach. We
consider a mixture of neutrons, protons, electrons and eventually
neutrinos in $\beta$ equilibrium inside each isolated WS cell assumed to
be electrically neutral. In the regime that we are interested in,
electrons are extremely relativistic and can be assumed to be uniformly
distributed in the cell. We solve the problem self-consistently at TF
level disregarding other particles, shell and pairing effects.

We consider first the outer crust of a neutron star where the WS cell
contains a nuclear cluster surrounded by a neutron gas with an average
density in the range $4.3 \times 10^{11}$ g/cm$^3 \lesssim \rho
\lesssim
2.0 \times 10^{14}$ g/cm$^3$.\cite{Neg73} In this scenario
temperature
is basically zero and there are no neutrinos trapped inside the WS
cell. Fig. 3 (left panel) displays the neutron and proton density
profiles obtained with our model assuming that the system with $Z=250$
is $\beta$-stable and the average density is $\rho=5.77\times 10^{-3}$
fm$^{-3}$ in the WS cell. This two conditions allow to determine the
radius of the WS cell which is $R=107$~fm and the number of nucleons
inside the cell: $A=29608$. The calculated energy per nucleon
is 1.846 MeV which is close to the HF ground-state energy per nucleon
at this average density.\cite{Neg73}
\begin{figure}
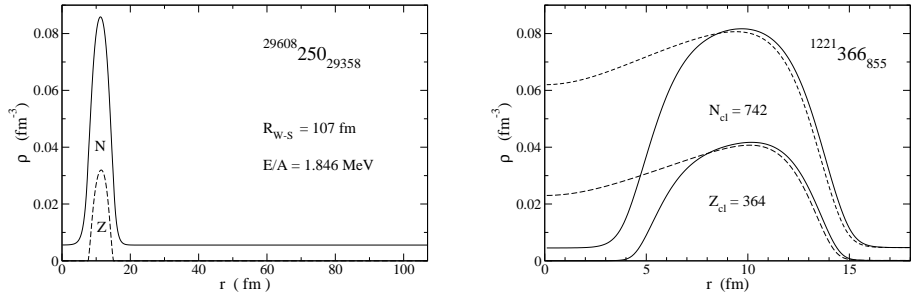

\begin{center}
\psfig{file=kazi_Bnstars.eps,width=5.6cm,clip=true}
\qquad
\psfig{file=kazi_Bsnova.eps,width=5.6cm,clip=true}
\caption{Neutron and proton density profiles in a Wigner-Seitz cell 
in the outer crust of a neutron star (left) and in conditions prevailing 
during the gravitational collapse of massive stars (right).}
\end{center}
\end{figure}

A different scenario concerns hot dense matter in conditions
prevailing during the gravitational collapse of massive stars. One key
ingredient for describing the collapse is just the EOS. In the
relatively low density regime $0.001$ fm$^{-3} \lesssim \rho \lesssim
0.05$ fm$^{-3}$ the most favored phase consists of nucleons
congregated into nuclear clusters which can be huge ($A \lesssim
1000$) and immersed in a free nucleon sea. In this case the WS cells
contain neutrons, protons, electron and trapped neutrinos, again under
the conditions of charge neutrality and $\beta$-equilibrium (for
details see Ref. \cite{Pi86} and references therein). 
We calculate one possible point of the EOS at a constant entropy per
nucleon $S/A=1 k_{\rm B}$, as far as the collapse is fairly
isentropic. The WS cell contains 855 neutrons and 366 protons
corresponding to an average density $\rho=0.05$~fm$^{-3}$ and a proton
concentration $Y_e=0.3$, which are typical conditions in this
scenario.\cite{Pi86} The neutron and proton density profiles obtained
by minimization of the {\it total} free energy $F$ (including
electrons and neutrinos) are displayed in Fig.~3 (right panel). We
obtain two possible solutions with almost the same free energy per
particle. One corresponds to a semi-bubble distribution ($F/A= 28.86$
MeV) and the other one to a true bubble embedded in a low density
fluid ($F/A= 28.73$ MeV).

From these simple estimates in stellar media we conclude that
spherical semi-bubbles and true bubbles could in principle exist in
the considered astrophysical context. Whether these bubbles
are stable against deformations demands further investigations.

\section{Toroidal structures}

Long-lived nuclear structures beyond the island of stability may be
reached if other topologies different from the sphere are considered.
Simulations using Boltzmann-Uehling-Uhlenbeck (BUU) transport
equations have shown that exotic shapes such as bubbles and tori could
be created in central heavy-ion collisions.\cite{Mor97} Some effort
has been addressed to find experimental evidences of the breakup of
toroidal distributions produced in heavy-ion collisions.\cite{TorExp}

Nuclei of toroidal shape were first analyzed by Wong in the framework
of the LDM plus Strutinsky shell corrections.\cite{Won73} It was
found that there exists a minimum of energy for such a type of shape
which, however, is usually unstable against multifragmentation. Within
mean-field theory using Skyrme forces the toroidal structure in
superheavy nuclei was predicted by Nazarewicz el al.\cite{Naz02}
More recently, axially symmetric toroidal superheavy nuclei with $Z
\ge$ 130 have been analyzed by one of us within the HFB theory with
the Gogny D1S force.\cite{War07} A global minimum of the binding
energy in toroidal distributions is obtained by performing constrained
HFB calculations with a relatively {\it large} and {\it negative} mass
quadrupole moment $Q_2$. It is also found that for mass numbers $A$
large enough, the superheavy toroidal nucleus becomes the ground
state and it can be as much as 200~MeV more bound than the
corresponding spherical nucleus.

We perform here deformed ETF calculations including
$\hbar^2$-corrections with the SkM$^*$ force to analyze
semiclassically toroidal nuclei. The Weizs\"acker term $\frac{1}{36}
\, f_{q} (\mbox{\boldmath $\nabla$} \rho_q)^2/\rho_q$ of the kinetic
energy density given in Eq.\ (\ref{eq2}) is multiplied by a factor
1.6 in such a way that the ETF surface energy calculated with the
SkM$^*$ interaction equals the HF value.\cite{Gar89}

\begin{figure}
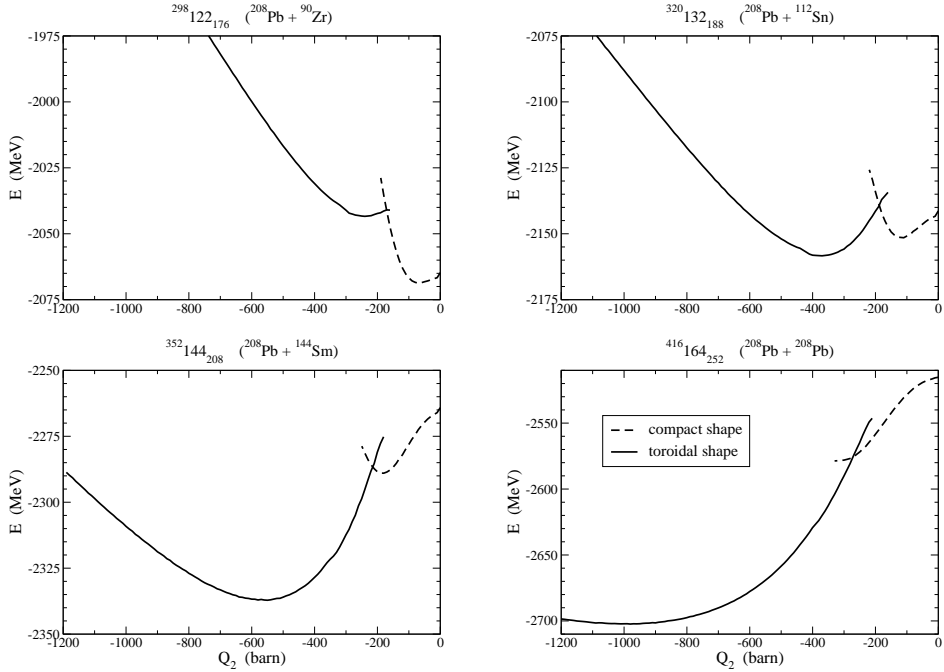

\begin{center}
\psfig{file=kazi_Pb_Zr_energy.eps,width=5.8cm}
\qquad
\psfig{file=kazi_Pb_Sn_energy.eps,width=5.8cm}\\
\psfig{file=kazi_Pb_Sm_energy.eps,width=5.8cm}
\qquad
\psfig{file=kazi_Pb_Pb_energy.eps,width=5.8cm}
\caption{The PES corresponding to $^{298}$122, $^{320}$132, $^{352}$144 and
$^{416}$164  systems for negative values of the mass quadrupole moment $Q_2$.}
\end{center}
\end{figure}

Fig. 4 displays the potential energy surface (PES) as a function of the
quadrupole mass moment $Q_2$ for systems with $^AZ =$ $^{298}122$, $^{320}$132,
$^{352}144$ and $^{416}$164. Such systems would hypothetically be created in 
$^{208}$Pb + $^{90}$Zr, $^{208}$Pb + $^{112}$Sn, $^{208}$Pb + $^{144}$Sm and
$^{208}$Pb + $^{208}$Pb collisions of stable nuclei, respectively. In all the
analyzed cases we see that the PES consists of two branches that cross
at a given value of $Q_2$. The crossing point is shifted to more negative values
of $Q_2$ when the mass number of the system increases. The dashed branches in
the Fig 4. correspond to a compact oblate distribution and the solid ones  to a
toroidal shape. These two types of nuclear distributions can be seen
in Fig.~5 where 2-dimensional plots of the self-consistent density of
the nucleus $^{416}$164 are displayed for the compact solution at
$Q_2=-250$ b and for the toroidal structure at quadrupole moments
$Q_2=-250$ b, $-650$ b and $-990$ b (ground state).

For the lightest system $^{298}$122, the ground state corresponds to a compact
oblate shape with comparatively small negative deformation. However, when $A$
and $Z$ increase, the minimum point of the toroidal branch decreases and it
becomes the ground state for the $^{320}$132 system, as well as for the heavier
systems. From Fig.~4 one also observes that when the mass and the charge of the
system increase the ground-state energy is more negative as compared with the
energy of the spherical solution ($Q_2=0$). For heavy systems the minimum
becomes more flat and its position is shifted to more negative values of the
quadrupole moment making the toroidal nucleus unstable against
multifragmentation. The PESs of toroidal nuclei calculated in the Skyrme ETF
method (without shell corrections) are in qualitative agreement with previous
findings in the mean-field HFB theory with the Gogny force, although there are
differences in the absolute values of the binding energies of the
system.\cite{War07}

\begin{figure}
\begin{center}
\psfig{file=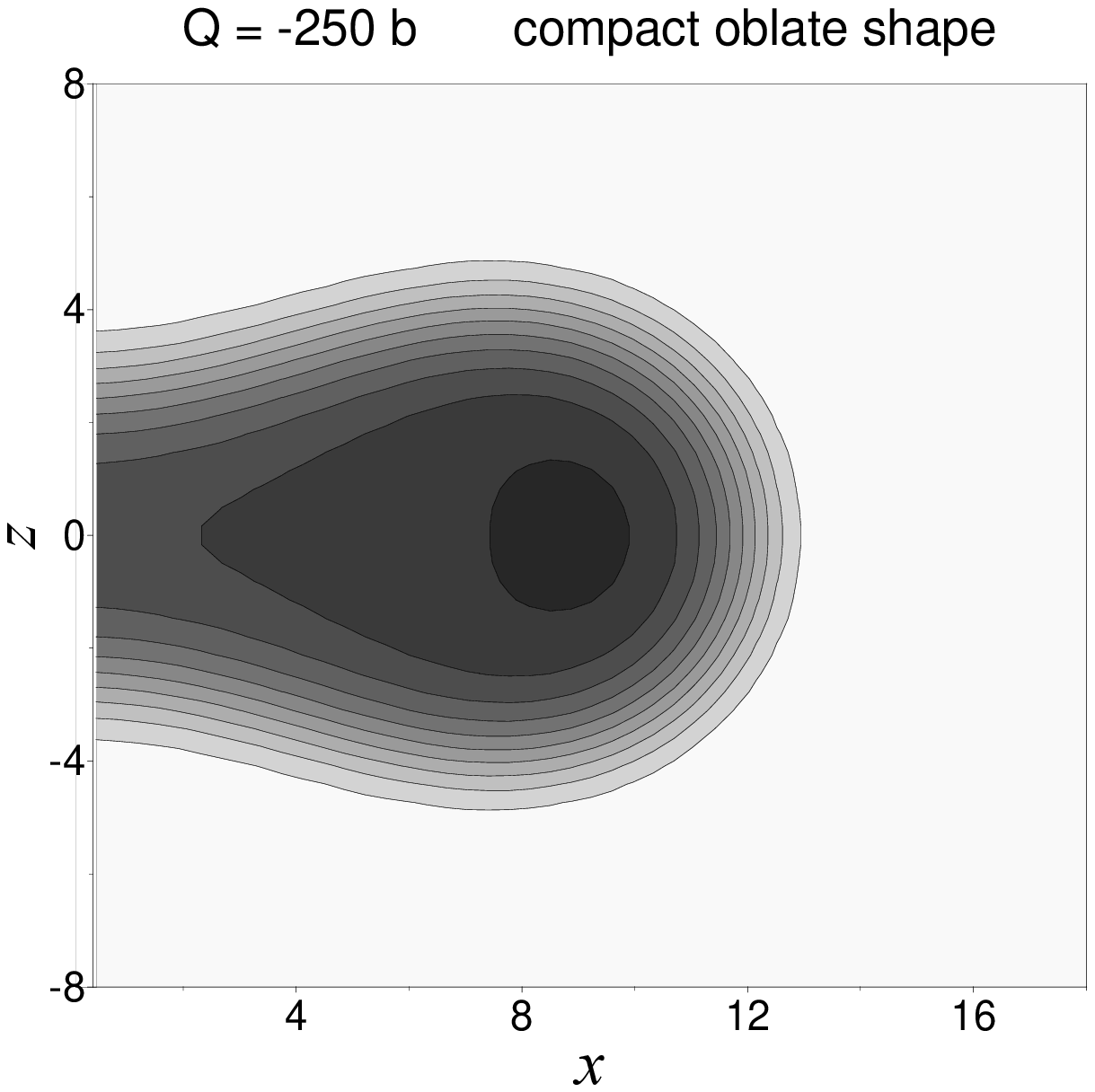,width=5.5cm}
\quad
\psfig{file=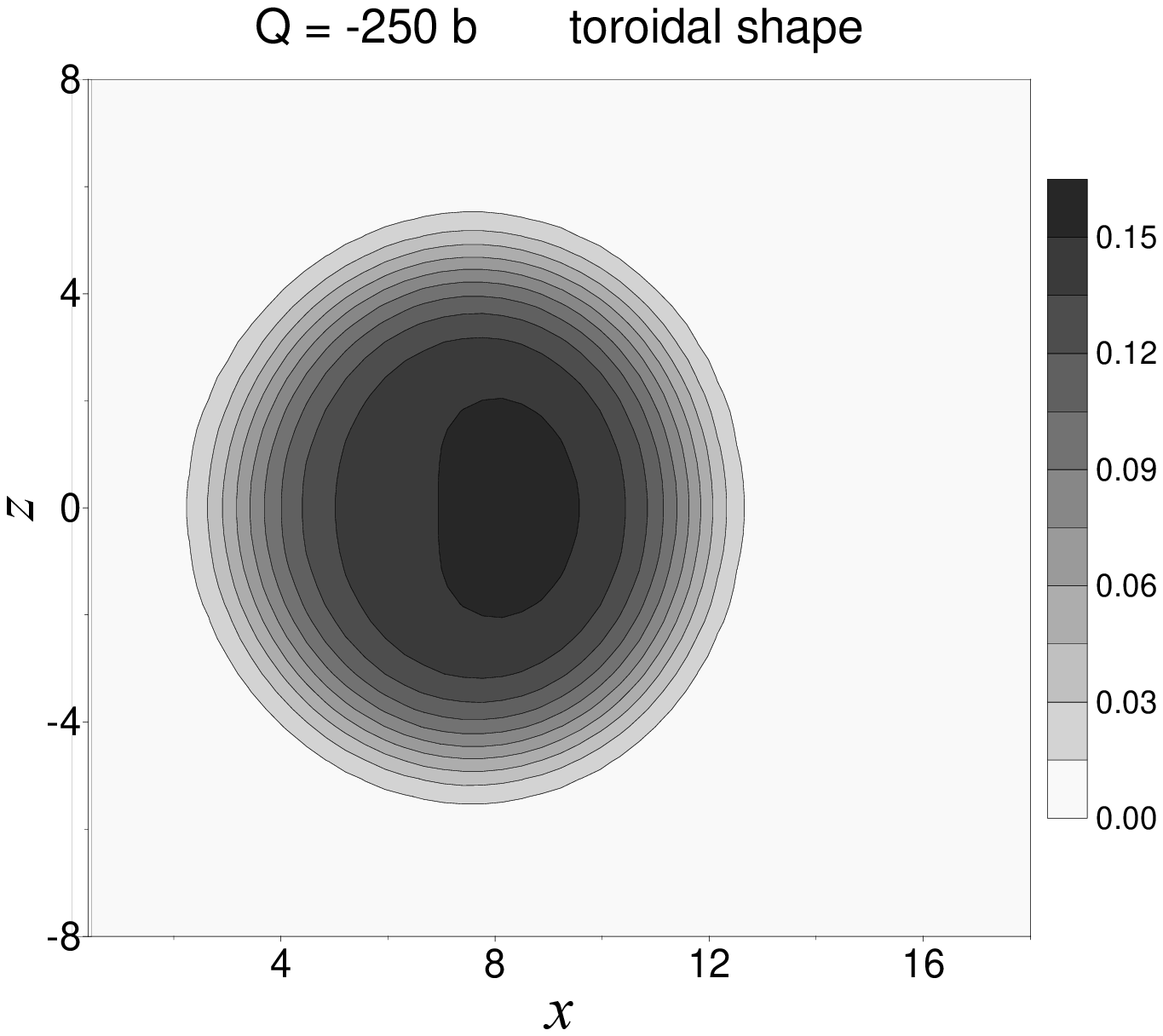,width=6.2cm}\\[2mm]
\psfig{file=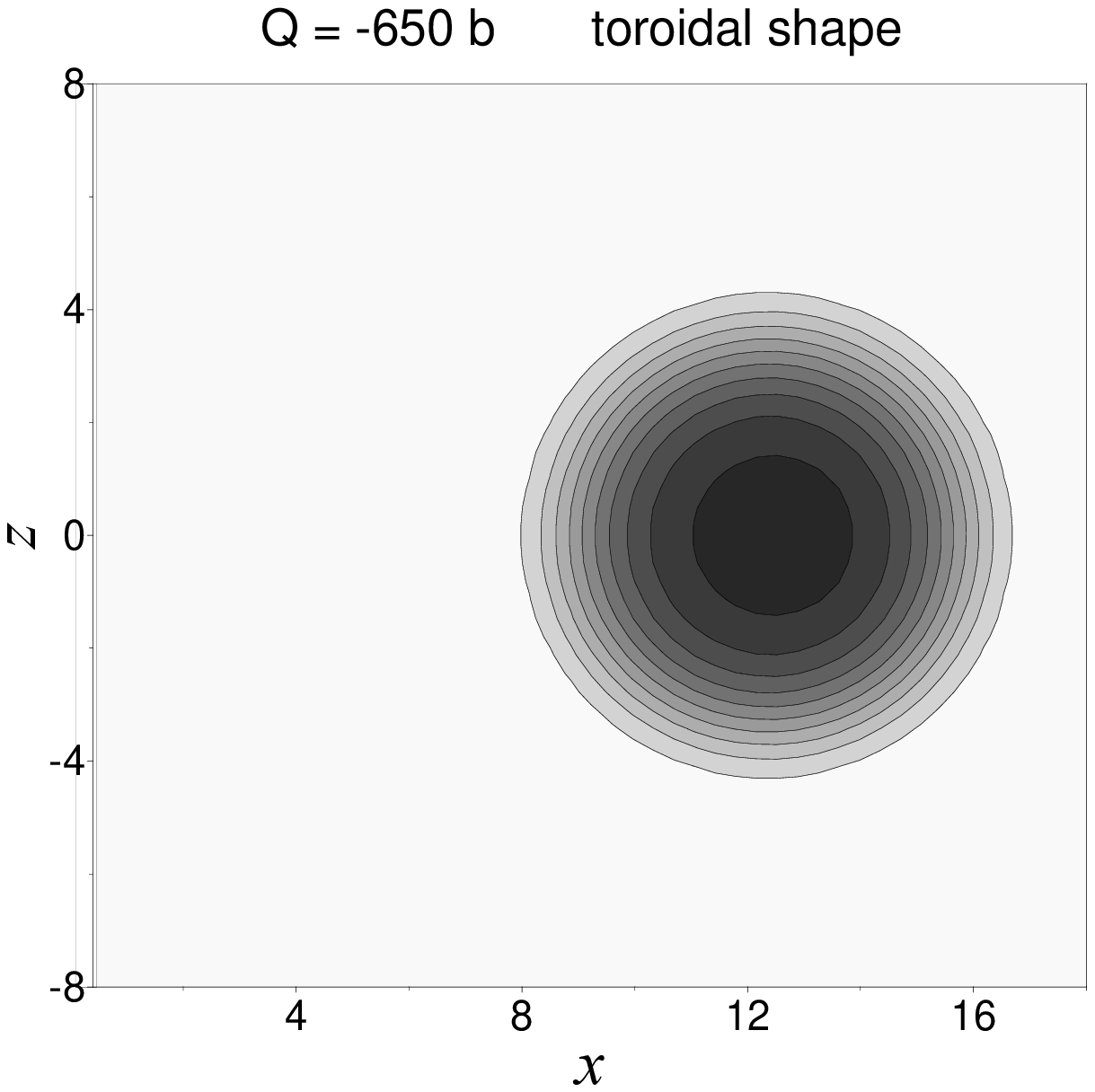,width=5.5cm}
\quad
\psfig{file=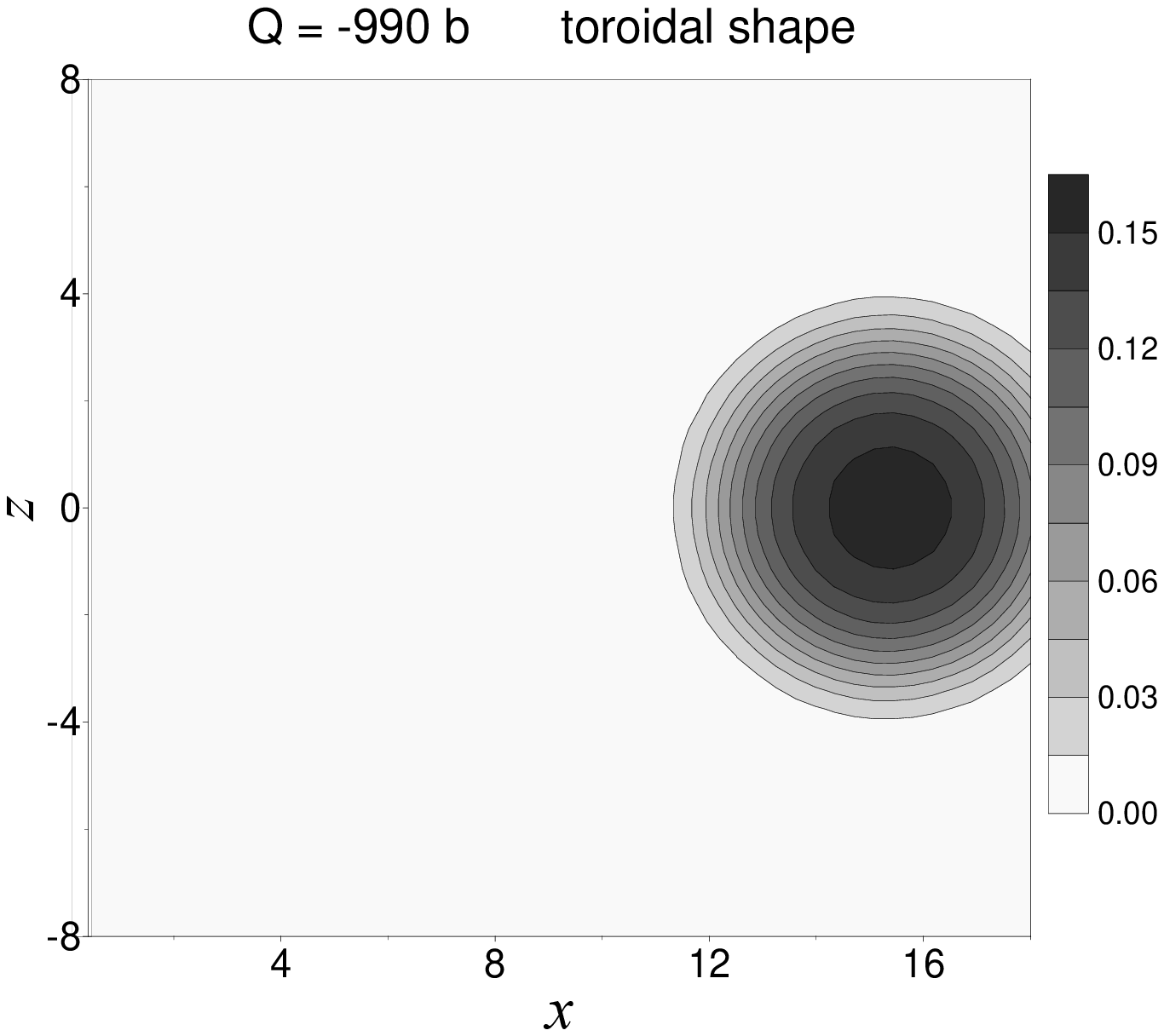,width=6.2cm}\\
\caption{Two-dimensional density plots in the $x z$ plane
for the nucleus $^{416}$164 at
several mass quadrupole moments $Q_2$. The configurations are axially
symmetric around the $z$ direction.}
\end{center}
\end{figure}

In a small range of quadrupole momenta around the crossing $Q_2$
values we find numerically two possible solutions of the EL equations
(\ref{eq5}) for the same value of $Q_2$. These two solutions have
close binding energies. One of them belongs to the compact branch and
the other one to the toroidal branch in Fig.~4. (This situation
resembles the one discussed previously for spherical bubble
distributions, where it is possible to find simultaneously
semi-bubbles and true bubbles for some values of mass and atomic
numbers.) In the case of the $^{416}$164 system, the two solutions are
found in the range $-350 \lesssim Q_2 \lesssim -200$ b (see Fig.~4).
Two-dimensional contour plots of the compact and toroidal density
distributions at $Q_2= -250$ b for $^{416}$164 are depicted in the
upper panels of Fig.~5.

It is interesting to inspect the modification of the calculated PES
induced by a finite temperature $T$ and by the rotation of the torus
around a radial axis with angular momentum $L_x$. The stability of hot
rotating toroidal structures was studied in the past by
Wong\cite{Wong78} in the framework of the LDM. In the left panel of
Fig.~6 we display our ETF results for the case of the $^{416}$164
system at $T=0$ and $T=1$ MeV, and when the system is rotating with
angular momentum $L_x= 150\,\hbar$. To simplify the figure, we use the
same line style for the compact and toroidal branches of the PES in
each studied case and do not draw the points in the crossing of the
two branches.

One observes that for small values of $T$ the excitation energy of the
$^{416}$164 nucleus is roughly independent of the deformation $Q_2$.
With a temperature $T= 1$~MeV the energy is shifted upwards about
$55-60$ MeV along the whole PES. This value can be seen as the
difference between the dashed and solid curves in the left panel of
Fig.~6. We treat rotation in the rigid-body approximation as described
in Ref.\ \cite{Cen93}. The excitation energy gained by the $^{416}$164
system under rotation is more important for more compact (i.e. less
deformed) distributions, as it can be expected from the reduction of
the rigid-body moment of inertia when $Q_2$ increases (i.e. is less
negative). In the present example, the energy gain of the system with
angular momentum $L_x= 150\,\hbar$ with respect to the non-rotating
system is around $30-35$ MeV in the compact branch of the PES, while
it decreases from about $20$ MeV at $Q_2=-350$~b to only $7$ MeV in
the region around the minimum of the PES.

It is to be noted that the PES of toroidal nuclei depends on the value
of the surface energy associated with the nuclear interaction. To
exemplify this, we display in the right panel of Fig.~6 the PES of the
$^{416}$164 system obtained multiplying the Weizs\"acker term in
Eq.\ (\ref{eq2}) by a factor of 2 (to reproduce the LDM fission
barrier of $^{240}$Pu)\cite{Gar89} instead of by the 1.6 factor used
in the previous calculations. It can be seen that in this case the
minimum exhibited by the PES is more marked and shifted towards more
compact (less negative $Q_2$) configurations. Consequently, it is
expected that nuclear forces with large surface energy make toroidal
nuclei more stable against multifragmentation.

 \begin{figure}
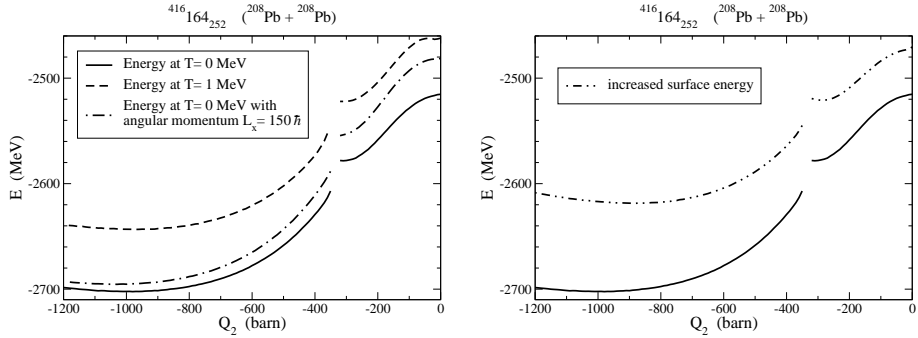

\begin{center}
\psfig{file=kazi_Pb_Pb_energy_TEMP.eps,width=5.8cm}
\quad
\psfig{file=kazi_Pb_Pb_energy_betakt2.eps,width=5.8cm}
\caption{Thermal and rotational (left) and surface energy (right) effects
on the PES of the $^{416}$164 nucleus as a function of the mass quadrupole moment
$Q_2$.} 
\end{center}
\end{figure}

\section{Summary and outlook}

In this work exotic nuclear shapes, namely bubbles and tori, containing several
hundreds of nucleons are analyzed  by means of the semiclassical extended
Thomas-Fermi method and the Skyrme force SkM$^*$. Calculations are performed by
solving fully self-consistently the variational Euler-Lagrange equations in
spherical and axial symmetries. The ETF method incorporates in a consistent way
the description of the nuclear surface and is free of shell effects. Although
shell effects are crucial for the stability against deformation of these huge
and exotic nuclear clusters, important information about the smooth variation
with the number of particles of some global properties can be obtained
from the semiclassical analysis.

As it happens in the HF calculations, for some neutron and proton
numbers two different solutions of the ETF equations, namely
semi-bubbles and true bubbles, are obtained. For a large enough number
of nucleons the true bubbles are the structure with minimal energy. We
have also explored thermal effects in spherical bubbles. In this
picture, spherical bubbles can exist at temperatures above the one for
which shell effects disappear but with a limiting temperature smaller
than for normal nuclei. Using a simplified model we have also explored
the possible existence of spherical bubbles in astrophysical
scenarios. From a semiclassical point of view it is found that
semi-bubbles and true bubbles could exist in the outer crust of
neutron stars and in hot and dense stellar matter in conditions
prevailing during the gravitational collapse of massive stars.

We have performed deformed Extended Thomas-Fermi calculations for
toroidal distributions of nucleons. Threshold values of the mass and
atomic numbers exist beyond which toroidal nuclei become the ground
state of the system. With increasing mass and charge the ground state
of toroidal nuclei is more deformed and more bound with respect to the
spherical solution, undergoing ultimately decay by multifragmentaion.
Effects of temperature, rotation and value of the surface energy on
the PES of tori have also been discussed.

The semiclassical study presented here is a step towards a consistent
description of bubbles and tori which has to include shell corrections
computed from the semiclassical single-particle potentials. In this
way it is expected to obtain a quantal description of these exotic
nuclear clusters that will allow to know separately the smoothly varying
and the oscillatory parts of some nuclear properties. The preliminary results
obtained in astrophysical scenarios are very encouraging. An important point is
to investigate the stability against deformation of bubbles embedded in a gas
containing nucleons, electrons and neutrinos.

\section*{Acknowledgments}
Work supported in part by Grants No.\ FIS2005-03142 from MEC (Spain)
and  FEDER, No.\ 2005SGR-00343 from DGR (Generalitat de Catalunya)
and No.\ N202 179 31/3920 from MNiSW (Poland).

\end{document}